\documentclass[article,submit,pdftex]{revtex4-2}%{Definitions/mdpi}
\usepackage[utf8]{inputenc}
\usepackage{amsmath}
\usepackage{amsfonts}
\usepackage{mathrsfs}
\usepackage{amssymb}
\usepackage{graphicx}
\usepackage{latexsym}
\usepackage{color}
\usepackage{booktabs}
\usepackage{dcolumn}
\usepackage{epsfig}
\usepackage{float}
\usepackage{multirow}
\usepackage{ulem}
\usepackage{svg}
\usepackage[mathcal]{euscript}
\usepackage{tikz}
\def\be{\begin{eqnarray}}
\def\ee{\end{eqnarray}}
\def\nn{\nonumber}
\begin{document}

\title{Lower dimensional black holes in nonlinear electrodynamics: causal structure and scalar perturbations}

\author{R. D. B. Fontana} % \orcidlink{0000-0002-8835-7447}}
\email[Email: ]{rodrigo.dalbosco@ufrgs.br}
\affiliation{Universidade Federal do Rio Grande do Sul, Campus Tramandaí-RS Estrada Tramandaí-Osório, Tramandaí CEP 95590-000, RS, Brazil}

\begin{abstract}

We study the charged black hole solutions of a 2+1 nonlinear electrodynamical theory with cosmological constant. Considered as a one-parameter group of theories (the exponent of the squared Maxwell tensor) the causal structure of all possible black holes is scrutinized. We analyze the singularity character that each theory delivers together with their horizons and the plausible limitations in the black hole charges. The investigation demonstrates a rich structure of three different groups of theories, according to the qualitative behavior of the singularity, horizons and limitations in the geometric charges. For such groups we study the effect of a scalar field propagating in the fixed black holes spacetime. All geometries analyzed were stable to such linear perturbations, these evolving as usual quasinormal spectra of the black holes that we calculate in different cases.
\end{abstract}
%
%\keyword{lower dimensional gravity; nonlinear electrodynamics; black hole perturbations; causal structure;} 
%
%

\maketitle

%%%%%%%%%%%%%%%%%%%%%%%%%%%%%%%%%%%%%%%%%%

%%%%%%%%%%%%%%%%%%%%%%%%%%%%%%%%%%%%%%%%%%
\section{Introduction}

Lower dimensional theories of curvature represent an interesting and extensive branch of research in actual days \cite{Deser:1983nh,Carlip_1998,García-Díaz_2017a}. As a toy model for testing boundary properties of gravity it correlates to physical features of two (or one) dimensional conformal field theory where the physics is generally better understood \cite{Arenas1,Arenas2}.

Since the pioneer works of Bañados {\it{et al}} \cite{btz1}, Jackiw \cite{Jackiw1} and Mann \cite{Mann:1991qp}, a plethora of studies of such theories came to light. In the outstanding work \cite{btz1}, although no graviton is found, the geometry possess a black hole as a possible solution to the curvature equations if the action includes a negative cosmological constant. The AdS boundary of the spacetime still presents the notable shape that allows for the investigation of the AdS/CFT correspondence.

The causal structure of the black holes in lower dimensional gravity is considerably affected by the matter background. The singularity character in a BTZ line-element, is different when charge, rotation or fluids are present \cite{deOliveira}. \textcolor{black}{We have a regular black hole at $r=0$ for a BTZ black hole with mass (see e. g. \cite{reg3d}), rotation and cosmological term and a curvature singularity if charge is present}. In the case of a Kiselev (2+1)-BTZ solution  \cite{kiselev1} we may also have lightlike singularities \cite{deOliveira}. 

The presence of horizons is also significantly affected for the existing matter in the action. In the linear theory, geometries with charge and/or angular momentum have two horizons (event and Cauchy), while in their absence, spacetimes with cosmological constant and fluids have only one horizon \cite{deOliveira}. We may further see, in the nonlinear electrodynamical regime we have a more intricate behavior depending on the specific theory.

After the announcement of \cite{btz1}, examples of black hole solutions in lower dimensional gravities are numerous \cite{Holst:1999tc, Cataldo:2000we, Emparan:2020znc, Koch1, Contreras:2019iwm} \textcolor{black}{(e. g. specifically in scalar tensor theories see \cite{st1,st2,st3,st4,st5,str6,str7,str8})} and their properties have been extensively studied \cite{Birmingham:2001dt, Birmingham_2002, Lee:1998pd, Dasgupta:1998jg, Rincon:2019zxk, oliveira22,rincon25}. One of those is the dynamical stability of the geometry to field propagations (and its quasinormal modes) that is the second issue we tackle in this manuscript. 

The perturbations of black holes represent an important theoretical development in the study of the linear stability of such objects whose first work dates back to 1957 \cite{Regge:1957td}. Since then, the theory instigated several progresses far beyond its scope remaining nowadays as a cornerstone in the measurement of gravitational waves \cite{LIGOScientific:2016aoc, LIGOScientific:2016sjg, LIGOScientific:2017bnn}. In that direction, the fundamental quasinormal modes for classical general relativity were broadly studied until the 80´s in consideration to different field perturbations \cite{Zerilli:1970wzz, Zerilli:1970se, Teukolsky:1972my, teuko74, leaver1, leaver2, wkb1, wkb2, wkb3, wkb4}. 

In the case of black holes in (2+1) dimensions works for the BTZ solution with mass, angular momentum were performed in \cite{Cardoso_2001, bir1} and even more recently with charge \cite{Fontana_2024, Fontana:2023dix}. Although a complete study considering the perturbations of a complete BTZ black hole with rotation and charge is still missing, interesting results of superradiance of such black hole were reported in \cite{eksw}.

In the nonlinear electromagnetic cases, specific perturbations of black holes with charge were considered in \cite{Aragon_2021, Gonzalez_2021}, based on a particular case of a range of theories developed in \cite{hendi10,hendi14} \textcolor{black}{and other geometric properties of those theories were investigated in e. g. \cite{nled0, nled1, nled2, nled3,nled4,nled5}}.

In this work we will be concerned with two main aspects of these theories: the causal structure (existence of horizons and presence of a curvature singularity) and the stability to linear scalar perturbation with the calculation of the quasinormal modes. Those are the subjects we investigate in the paper. 

In the next section we present the theory and possible black holes resulting of that, analyzing its limits for different charges inspecting their effect in the possible horizons. In section \ref{sec3} we characterize the dynamics of a scalar field propagating in such background, developing the numerical technique we use to investigate the perturbation problem. Our results on the stability issue and quasinormal modes are delivered in section \ref{sec4} which is followed by our final remarks on section \ref{sec5}.

\section{Black holes in nonlinear electrodynamics}\label{sec2}

We start by considering the 2+1 theory that brings the possible background solutions given by \citep{hendi10,hendi14}, 
\be
\label{e1}
S = \int_{\partial \mathcal{M}} d^3x \sqrt{-g} \big( R-2\Lambda + (k\mathcal{F})^\sigma \big),
\ee
in which $\mathcal{F}= F_{\mu \nu}F^{\mu \nu}$. \textcolor{black}{Here $k$ is an electromagnetic constant that allows for the rescaling of the curvature part of 3-dimensional actions in its curvature term, considering its coupling constant as 1 (firstly proposed in \cite{btz1}). We may incorporate such constant in the definition of electromagnetic charge when defining the electromagnetic potential.} 

The resulting motion equations of the above action define the geometry background of our theory and the equations of curvature and Maxwell fields are written as
\be
\label{e2}
G_{\mu \nu} + \Lambda g_{\mu \nu} & = & T_{\mu \nu} \\
\label{e3}
\partial_\mu \Big( \sqrt{-g} F^{\mu \nu} (k\mathcal{F})^{\sigma-1} \Big) & = & 0,
\ee
in which the energy-momentum tensor is determined \textcolor{black}{by the} electromagnetic field, defined as
\be
\label{e4}
T_{\mu \nu} = 2\sigma k F_{\mu \alpha}F^{\alpha}_\nu (k \mathcal{F})^{\sigma -1} - \frac{1}{2}g_{\mu \nu} (k \mathcal{F})^{\sigma}.
\ee
Now, considering an spherically symmetric background Ansatz, 
\be
\label{e5}
ds^2 = -f(r)dt^2 + \frac{1}{f(r)}dr^2 + r^2 d \varphi^2
\ee
the usual Maxwell field of those spacetimes is chosen to be a general vector potential written in terms of the $r$ coordinate, $\mathscr{A} = A_\mu dx^\mu = A(r) dt$ bringing $\mathcal{F}= -2 (\partial_r A)^2$. The Maxwell tensor has only two non-zero components, $F_{tr} = -F_{rt} = \partial_r A$ rendering all three Maxwell equations as the same one, 
\be
\label{e6}
\frac{\partial A}{\partial {r}} + (2\sigma -1)r\frac{\partial^2 A}{\partial {r}^2}  =0.
\ee  
Here, for $\sigma \neq 1$ we have a simple solution,
\be
\label{e7}
A(r) = Q r^{\frac{2\sigma - 2}{2\sigma -1}}
\ee
(the case $\sigma = 1$, the linear electrodynamics, was already extensively studied in the literature and we will not analyze it in this work). Finally with the solution of (\ref{e7}) all three non-trivial curvature equations are linearly the same and can be put into the form
\be
\label{e8}
\frac{\partial f}{\partial r} + 2r\Lambda + \frac{(\sigma -1)^{2\sigma}}{(2\sigma -1)^{2\sigma-1}}(-8kQ^2)^\sigma r^{\frac{1}{1-2\sigma}} = 0
\ee
after what $f$ is written as
\be
\label{e9}
f(r) = -M + \frac{r^2}{\mathfrak{L}^2} + \mathrm{s}q^2 r^{\frac{2\sigma -2 }{2\sigma -1}}
\ee
in which $\Lambda = -\frac{1}{\mathfrak{L}^2}$, $M$ is an integration constant associated with the mass parameter, $q$ the redefined charge parameter and s the signal of the expression $1-\sigma$. 

Interestingly enough the lapse function (\ref{e9}) gives rises to a rich causal structure related to range of $\sigma >0$. For instance as we may see, different values of $\sigma$ produce black holes with one, two or no event horizon. \textcolor{black}{The thermodynamical aspects of those geometry was studied in \cite{hendi14} with the usual first law and entropy similar to that of the BTZ black hole with charge. For instance the event horizon Hawking temperature considering (\ref{e9}) is written as
\be
\label{e9b}
T_+ = \frac{2r_+}{4\pi \mathfrak{L}^2} +  \frac{\mathrm{s}q^2r_+^{\frac{1}{1-2\sigma}}}{4\pi \mathfrak{L}^2}\left( \frac{2\sigma -2}{2\sigma -1} \right).
\ee}

The limits on the $\sigma$-parameter and the subsequent causal structure generated is the focus of our attention in the next paragraphs.\\

{\it Theories in the scope of $0<\sigma < 1/2$.} 
\\

\begin{figure}[H]
\includegraphics[width=10 cm]{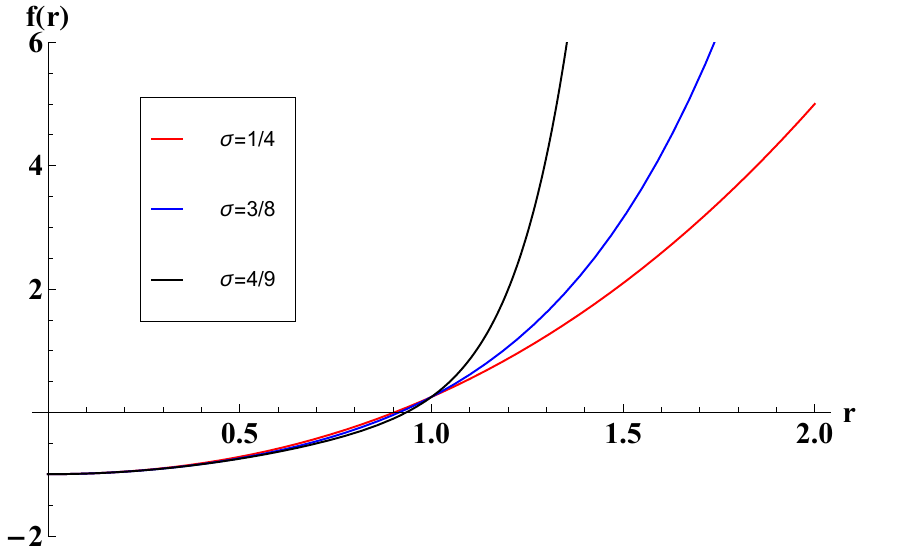}
\caption{The lapse function of black holes with $M=\mathfrak{L}=2q=1$ in the scope $0<\sigma < 1/2$.}
\label{fig1}
\end{figure}

In the prescribed range, $s=1$ and we define $\Sigma \equiv \frac{2\sigma -2 }{2\sigma - 1} $, a positive exponent, with $\Sigma  > 2$. In the asymptotic limits we have 
\be
\label{e10}
f\Big|_{r \rightarrow 0} \rightarrow -M, \hspace{2.0cm} f\Big|_{r \rightarrow \infty} \rightarrow \infty
\ee
with an inflection point $r_i$ to the relation $\partial_r f \Big|_{r_i} = 0$ expressed as
\be
\label{e11}
r_i = \left( \frac{1-\sigma}{2\sigma -1} q^2\mathfrak{L}^2  \right)^{\frac{2\sigma -1}{2\sigma }}.
\ee
Since the parentheses is negative, we have either $r_i <0$ or $r_i \in \mathbf{C}$. Considering the curvature scalars,
\be
\nn
\mathcal{R} & = & 6\Lambda - \frac{q^2\Sigma (1+\Sigma ) }{r^{\Sigma -2}}\\
\nn
\mathcal{R}_{\mu \nu} \mathcal{R}^{\mu \nu} & = & - 12\Lambda^2 +\frac{4\Lambda q^2\Sigma (1+\Sigma ) }{r^{2-\Sigma }} + \frac{q^4 \Sigma^2 (2+\Sigma^2 ) }{2r^{4-2\Sigma }}\\
\mathcal{R}_{\mu \nu \alpha \beta} \mathcal{R}^{\mu \nu \alpha \beta} & = & - 12\Lambda^2 +\frac{4\Lambda q^2\Sigma (1+\Sigma ) }{r^{2-\Sigma }} + \frac{q^4 \Sigma^2 (3+\Sigma^2 -2\Sigma ) }{2r^{4-2\Sigma }}
\label{e12}
\ee
it is possible to extend the black hole solution beyond $r=0$ if $\Sigma \geq 2$, which is the case. However for $r_i$ to be real, we have an extra condition,
\be
\label{e13}
 \frac{2\sigma -1}{2\sigma} = \frac{n_i}{2n_j+1}
\ee
 with $n_{i,j} \in \mathbf{N}$. Equation (\ref{e13}) can not be fulfilled for any pair $n_{i,j}$ that respects the scope of $\sigma$ we are restrict to, thus not allowing any real inflection point. The spacetime presented in that scope have no bounds for the constants $M,q$ and $\mathfrak{L}$ possessing always only one \textcolor{black}{horizon (the event horizon)}. Its lapse function is exemplified in figure (\ref{fig1}). Remarkably, the geometry is regular at $r=0$ which is an opposite feature of the traditional charged BTZ object: when $\sigma = 1$ the logarithm term at $f$ renders a curvature singularity at the origin.
%\unskip

As a last comment we must emphasize another peculiar behavior of the black holes in this particular range for $\sigma$: the geometry does not have a Cauchy (\textcolor{black}{inner}) horizon usually seen in charged geometries, being the event horizon the only one splitting the manifold in blocks.\footnote{It is still worth mentioning the singularity of the chargeless \textcolor{black}{BTZ geometry without rotation as being associated with an angular deficit in $\varphi$ if the line-element is considered isometric to the pure AdS geometry.}}
\\

{\it Theories in the scope of $1/2<\sigma < 1$.} 
\\

Here s is also positive and the range of $\sigma$ allows for only (any) negative values of $\Sigma$. According to (\ref{e12}) we can no longer extend the Manifold beyond $r=0$ as a curvature singularity is present in such point. The asymptotic regions of $f$ are given by
\be
\label{e14}
f\Big|_{r \rightarrow 0} \rightarrow \infty, \hspace{2.0cm} f\Big|_{r \rightarrow \infty} \rightarrow \infty
\ee
thus claiming the positive real solution of $r_i$ written in (\ref{e11}). In such case the signal of $f\Big|_{r_i}$ settles the number o event horizons of the solution: we may have two positive and real solutions for $f\Big|_{r_{+,-}} = 0$ representing Cauchy ($r_-$) and event ($r_+$) horizons or no solution at all. Since in our case $\Sigma \leq 0$, the curvature scalars diverge at $r \rightarrow 0$ in the last case (no horizons) we have a curvature naked singularity. We can avoid that ill-defined spacetime (\textcolor{black}{with a naked singularity}) by limiting $q$. 

Here a threshold for the charge is achieved in the equation $f\Big|_{r_i} = 0$ - representing solutions with a maximum charge $q=q_m$
\be
\label{e15}
-M + \frac{1}{\mathfrak{L}^2} \left( \frac{1-\sigma}{2\sigma -1} q_m^2\mathfrak{L}^2  \right)^{\frac{2\sigma -1}{\sigma }} + q_m^2\left( \frac{1-\sigma}{2\sigma -1} q_m^2\mathfrak{L}^2  \right)^{\frac{2\sigma -2}{2\sigma }}=0
\ee
that have both the Cauchy and event horizons at the same point. As a consequence whenever $q>q_m$ we have a naked singularity and for $q\leq q_m$ the geometry possesses the usual two horizons. We can simplify the above equation by redefining $\mathscr{Q} =   \left( \frac{- \Sigma q_m^2 \mathfrak{L}^2 }{2} \right)^{\frac{2\sigma - 1}{\sigma}}$  obtaining
\be
\label{e16}
\mathscr{Q} = \frac{\Sigma}{\Sigma -2}M\mathfrak{L}^2
\ee
or explicitly
\be
\label{e17}
q_m = \left( \frac{1-\sigma}{\sigma} M\mathfrak{L}^2  \right)^{\frac{\sigma}{4\sigma -2}}\sqrt{\frac{2\sigma -1}{(1-\sigma) \mathfrak{L}^2}}.
\ee
\begin{figure}[H]
%\isPreprints{\centering}{}{ }% Only used for preprints
%\begin{adjustwidth}{-\extralength}{0cm}
\centering
\includegraphics[width=0.48\textwidth]{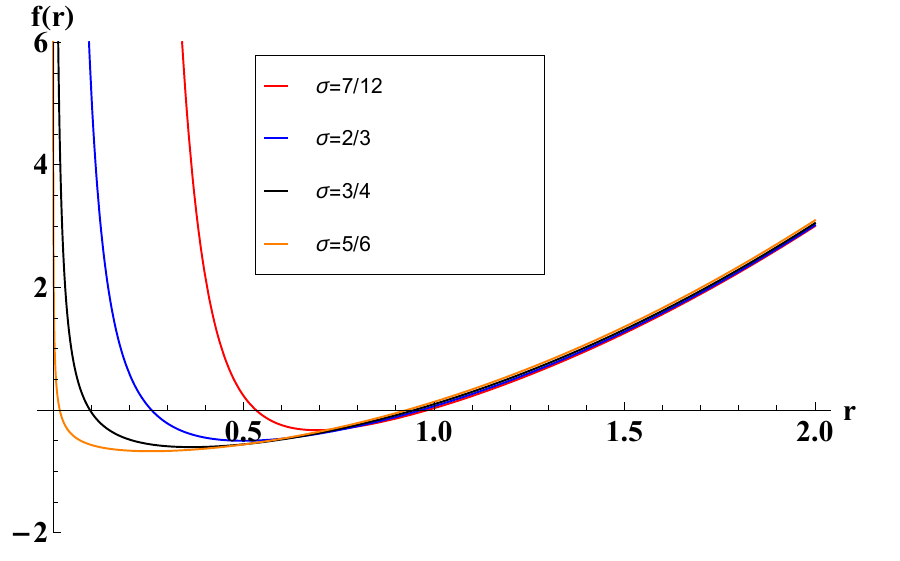}
\includegraphics[width=0.48\textwidth]{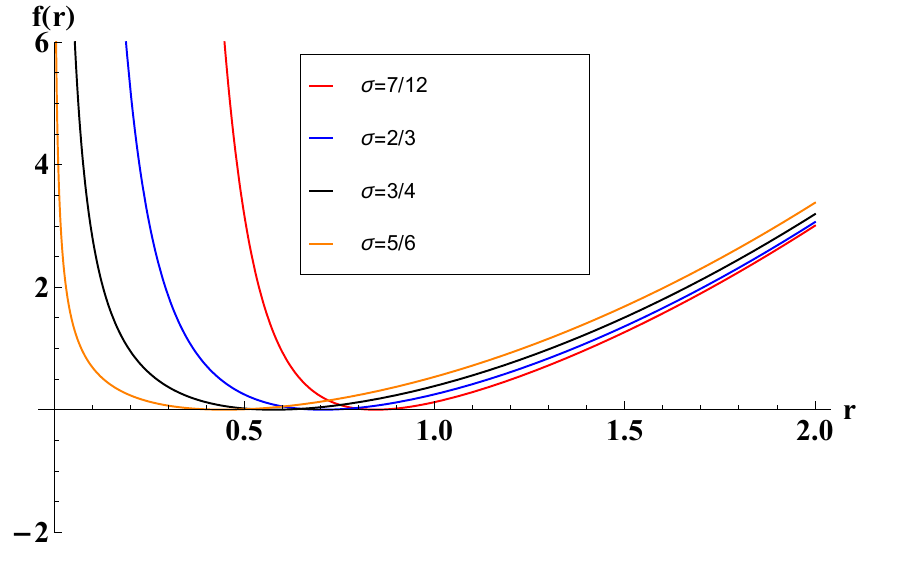}
\includegraphics[width=0.5\textwidth]{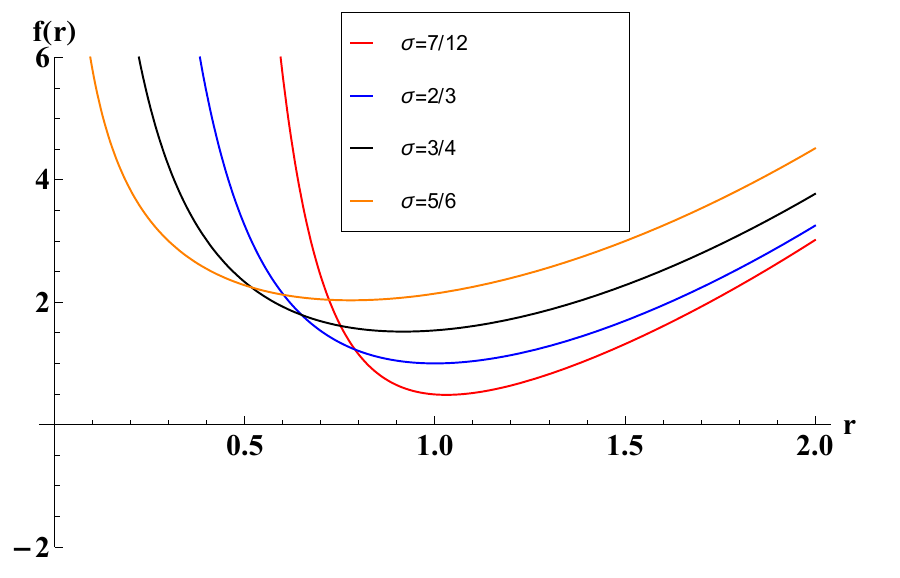}
%\isPreprints{\centering}{}{ }
%\end{adjustwidth}
\caption{The lapse function profile of black holes with $M=\mathfrak{L}=1$ in the scope $1/2<\sigma < 1$ with charges $q_m/2$ (left), $q_m$ (middle) and $2q_m$ (right).}
\label{fig2}
\end{figure}   
\unskip
In the present manuscript for the above $\sigma$-range we will considerthe propagation of a scalar field only in well-defined pacetimes with two horizons, thus within $q \leq q_m$, preventing the presence of a naked singularity. In figure (\ref{fig2}) we display three different panels representing the qualitative behaviors of $f(r)$.
\\

{\it Theories in the scope of $\sigma > 1$.} 
\\

\begin{figure}[H]
%\isPreprints{\centering}{} % Only used for preprints
\includegraphics[width=10 cm]{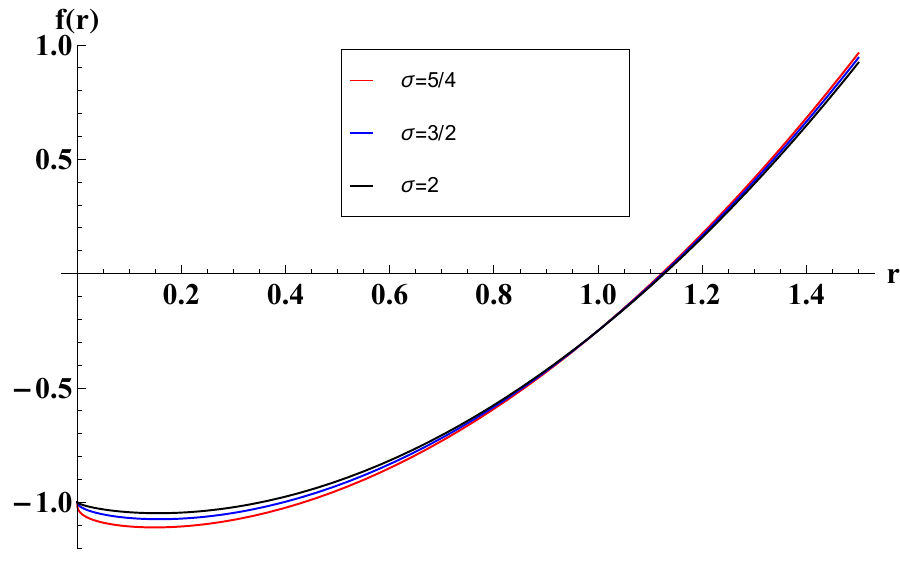}
\caption{The lapse function of black holes with $M=\mathfrak{L}=2q=1$ in the scope $\sigma > 1$.}
\label{fig3}
\end{figure}

In the last range for $\sigma$ we have s=-1 and $0<\Sigma < 1$. In this scope we also can not extend the Manifold beyond $r=0$ as a curvature singularity is present in such point. The asymptotic regions of $f$ read
\be
\label{e18}
f\Big|_{r \rightarrow 0} \rightarrow -M, \hspace{2.0cm} f\Big|_{r \rightarrow \infty} \rightarrow \infty .
\ee
Here we have the same inflection point defined in (\ref{e11}) in which $f\Big|_{r_i} < 0$. After $f(r_i)$, $f$ is a monotonically growing function defining exactly one solution to the equation $f(r)=0$, the event horizon. Again, we see no second horizon despite the presence of electromagnetic charge, what makes the spacetime particularly peculiar. For that kind of theories we see no bounds in the geometry constants $M, \Lambda$ and $q$ (no naked singularity is present whatever their values). A last figure of the lapse function in the last range is deployed in (\ref{fig3}).

Once the causal structure with the possible limits on the geometry constants is settled we are in position to study the dynamical stability of the black holes to small field perturbations. In this article we study the scalar field whose equation and methods are described in the next section.

\unskip

%* Scalar perturbation equations (1)
%* Results (5)
%* Final Remarks (1/2)
%* Introduction (2)
%* Abstract (1/2)

\section{Black hole dynamics: scalar field motion equation}\label{sec3}

The starting point for the studying the scalar perturbation in black holes backgrounds as those above described is the Klein-Gordon relation, 
\be
\label{e19}
\Box \Phi = \frac{1}{\sqrt{-g}}\partial_\mu \Big( g^{\mu \nu} \sqrt{-g} \partial_\nu \Phi \Big) = 0,
\ee
obtained through the variation of the scalar field matter action, $\int_{\mathcal{M}}d^3x \sqrt{-g}(\partial_\mu \Phi \partial^\mu \Phi)$. To develop (\ref{e19}) we choose the line-element (\ref{e5}) in the null ($dv = dt + f^{-1}dr$) and radial coordinates, bringing it to 
\be
\label{e20}
ds^2 = -f dv^2 +2dv dr + r^2 d\varphi^2.
\ee
Here the Klein-Gordon equation in such geometry is considerably simplified with a proper Ansatz for $\Phi$,
\be
\label{e21}
\Phi = \frac{\Psi (r)}{\sqrt{r}} e^{-i\omega v} e^{-i k \varphi}.
\ee 
with $k$ the angular momentum of the field, $k=0,1,2,\cdots$ and $\omega$ the harmonical dependence on $v$ also known as the quasinormal frequency.  The scalar equation turns to
\be
\label{e22}
f\partial_r \Big( f \partial_r \Psi \Big) - 2i\omega f \partial_r \Psi - V \Psi = 0,
\ee
$V$ being the effective potential,
\be
\label{e23}
V= f\left( \frac{k^2}{r^2}-\frac{f}{4r^2} + \frac{\partial_r f }{2r} \right).
\ee
The solution of (\ref{e22}) can be achieved with several numerical methods available in specific literature \cite{wkb1,wkb2,wkb3,wkb4,wkb5,wkb6,wkb7,leaver1,leaver2,Gundlach_1994, hh2000, jansen17}. We choose to work in the Frobenius method developed in \cite{hh2000}. For that, equation (\ref{e22}) must be written in the coordinate $x=1/r$,
\be
\label{e24}
0 = F^2x^4\partial_{xx} \Psi + Fx^2(Fx-F_{r}+2i\omega )\partial_{x} \Psi + \mathcal{V}\Psi \equiv s(x) \partial_{xx} \Psi + \tau (x) \partial_{x} \Psi + u(x) \Psi ,
\ee
 in which $F=f(r)\Big|_{r\rightarrow 1/x}$, $\mathcal{V}=V\Big|_{r\rightarrow 1/x}$ and  $F_{r} = \partial_r f(r)\Big| _{r\rightarrow 1/x}$. To solve (\ref{e24}) we expand each function $s, \tau$ and $u$ around $x_+=1/r_+$, 
\be
\label{e25}
s(x) = \sum_{n=0}s_n (x-x_+)^n \\
\label{e26}
\tau (x) = \sum_{n=0}\tau_n (x-x_+)^n \\
\label{e27}
u(x) = \sum_{n=0}u_n (x-x_+)^n,
\ee
and perform a similar expansion for $\Psi$,
\be
\label{e28}
\Psi = \sum_{n=0}a_n (x-x_+)^{n+\alpha}.
\ee 
We balance the perturbation equation order by order matching each coefficient in the expansion. To leading order two values of $\alpha$ are possible, $\alpha = 0$ and $\alpha = i\omega /\partial_rf(r_+)$. The last one represents an outgoing wave from the event horizon (classically irrelevant) and $\alpha = 0$ an ongoing front wave that we must consider for a physically relevant boundary condition. To higher order we have to satisfy the recurrence relation,
\be
\label{e29}
a_n = \frac{1}{n(n-1)s_2 + n\tau_1}\sum_{k=0}^{n-1}\Big( k(k-1)s_{n+2-k} + k\tau_{n+1-k} + u_{n-k}, \Big)a_k
\ee
and finally apply the condition at $r\rightarrow \infty$ that the field maintain its physical relevance, that is $\Psi|_{x=0} = 0$. This last relation is the one we can implement numerically truncating the series in a specific number of terms $N$,
\be
\label{e30}
\sum_{n=0}^N a_n(-x_+)^n = 0
\ee
checking the convergence with a higher $N_1>N$ afterwards. The quasinormal spectrum is achievable trough the above method and that are the results we report in the next section.  

%%%%%%%%%%%%%%%%%%%%%%%%%%%%%%%%%%%%%%%%%%
\section{Scalar field quasinormal modes}\label{sec4}

As a first step in our calculations with the scalar field perturbations, we aim to reproduce the results presented e. g. in \cite{Aragon_2021} when $\sigma = 3/4$. We accomplish that, with a precision higher than $0.1\%$. Particularly in the case of $M=\mathfrak{L}=1$ presented in tables 2 and 3, we obtained, e. g. $\omega_i = -1.83761i$ for $r_+=0.993$ as the highest deviation (cca $0.12\%$).  

The solutions for the linear theory $\sigma = 1$ - as mentioned in the section \ref{sec2} - were broadly studied in the literature since the original paper of 1992 and its companions \cite{btz1, Carlip_1995,Martinez_2000}. Particularly the scalar perturbations were considered in \cite{Fontana_2024, Fontana:2023dix} with their associated stability issues. For such reason we will not treat that solutions in the present work. 

\begin{figure}[H]
%\isPreprints{\centering}{} % Only used for preprints
\begin{center}
\includegraphics[width=0.45\textwidth]{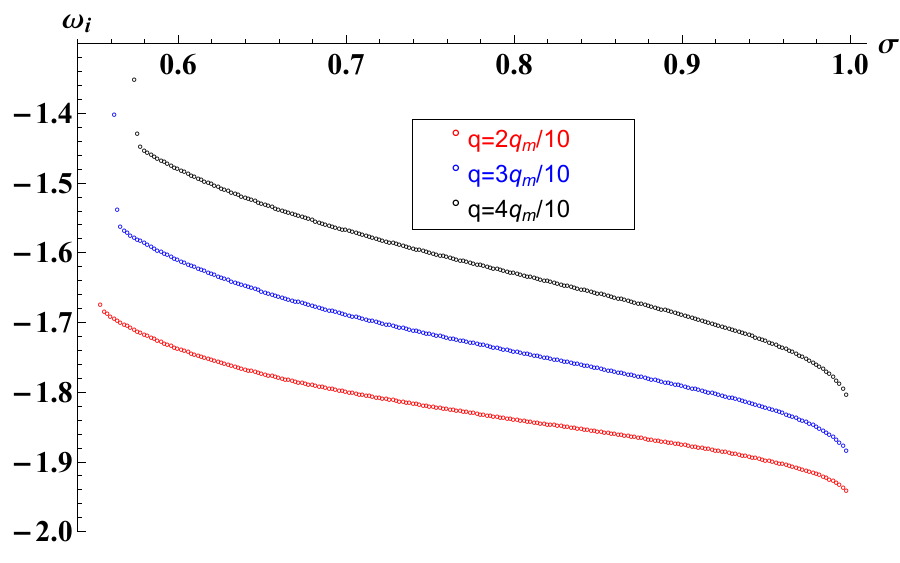}
\includegraphics[width=0.45\textwidth]{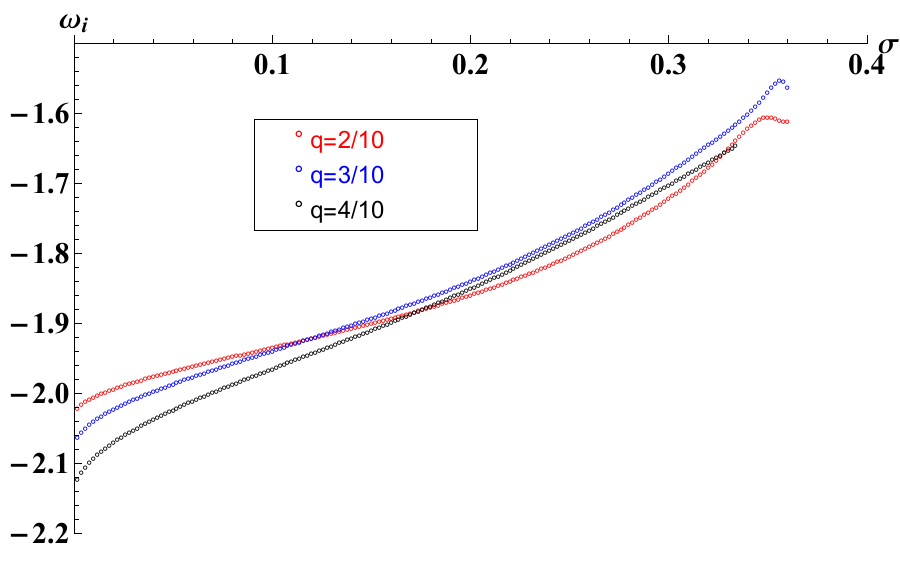}
\end{center}
\caption{The fundamental quasinormal modes of black holes in nonlinear electrodynamics. All the solutions are purely imaginary with no angular momentum ($k=0$). The geometry parameters are $M=\mathfrak{L}=1$.}
\label{fig4}
\end{figure}

After developing the Frobenius method numerically we employ it with a series expansion of 45 terms, checking it afterwards with a 60-tem series. The results have a very good convergence within less than $0.01\%$ of deviation. In figure {\ref{fig4}} we show the quansinormal modes for theories with $\sigma < 1$ considering multiple values of $q$.
%\begin{adjustwidth}{-\extralength}{0cm}
%\end{adjustwidth}
As displayed in the figure, all the fundamental frequencies are purely imaginary in the case of a massless field with no angular momentum ($k=0$). That is also the qualitative result found in other charged BTZ black holes as shown in \cite{Aragon_2021, Fontana_2024}.

 We emphasize two interesting features for the set of quasinormal modes we delivered in the panels. First of all, in theories with two horizons, we can see that an increase in the charge produces smaller values of $|\omega_i|$. Such behavior is qualitative the same as that found in \cite{Aragon_2021, Fontana_2024} although in the linear theory, the variation is much more pronounced. Second, in such theories, the increment in $\sigma$ diminishes $\omega_i$ such that, closer values to $\sigma = 1/2$ varies $\omega_i$ very abruptly. (This represents also the limit of convergence of the presented method).

If we consider small $\sigma$´s and one horizon theories, the behavior is qualitative the opposite as displayed in the right panel of \ref{fig4}. The smaller values of $\sigma$ present the higher $|\omega_i|$ while an increase in the coefficient diminishes the correspondent fundamental quasinormal mode. 

The peculiar structure of the modes demonstrated in both panels is an interesting novelty that locates $\sigma = 1/2$ as the special point of the series. Unfortunately the methods available do not allow the calculation of such theories in reason of the presence of $r^{\pm \infty}$ in the lapse function (for such reason any expansion near $\sigma = 1/2$ colapses the numerical solution).

We still pinpoint the departure value of quasinormal modes in both panels as $-2r_+i \sim -2i$ for the series of theories presented. That is in accordance to the the special cases when $\sigma \rightarrow 1$ (that would be the classical BTZ black hole without charge, and cosmological constant rescaled) and when $\sigma \rightarrow 0$ (also the original black hole without charge, considering a rescale in the mass term). 

In the range $0 < \sigma < 1/2$ as we have no bound for the spacetime charge we considered charges with higher values (and smaller $r_+$, $M=\mathfrak{L}=1$), generally forbidden in charged black holes. 
The results are deployed in table \ref{t1} and show that the oscillations are very sensitive to $q$ presenting a counterintuitive behavior: as we decrease $r_+$ - the area of the black hole, (increasing $q$) the absolute value of quasinormal modes increases (in contrast to the behavior reported in \cite{hh2000}). 

It is interesting to notice the stating point of $\omega_i$ for $\sigma = 0.01$, as $-5.8831i$. In the case $\sigma =0$ the frequency would be $-2i\sqrt{\frac{M(1+q^2\mathfrak{L}^2)}{\mathfrak{L}^2}} \sim -6.325i$ \cite{bir1}. With our methods for $\sigma = 10^{-6}$ we obtained $\omega_i \sim -6.330i$ and $\sigma =  10^{-7}$ brings $\omega_i \sim -6.327i$.
% Notice the temperature of the hole as presented by Heidi-2010 in (42)! That seems to couple with that stringent behavior. 

\begin{table}[h]
  \centering
 \caption{\color{black} Quasinormal modes for theories with $0<\sigma < 1/2$, high charge and zero angular momentum. The geometry parameters read $M=\mathfrak{L}=q/3=1$.}
%\addtolength\tabcolsep{6pt}
\setlength{\tabcolsep}{12pt} % Default value: 6pt
   \begin{tabular}{cccccccc}
%\vspace{0.2cm}
    \hline    \hline
%     \hline
\vspace{0.2cm}
	$\sigma$  &	$\omega_i$ &	$\sigma$  &	$\omega_i$ & $\sigma$  &	$\omega_i$ & $\sigma$  &	$\omega_i$  \\
%\vspace{0.2cm}
\hline \hline
%\vspace{0.2cm}
0.01	&	-5.8831	&	0.11	&	-4.8112	&	0.21	&	-4.2616	&	0.31	&	-3.9206	\\
0.02	&	-5.6928	&	0.12	&	-4.7443	&	0.22	&	-4.2187	&	0.32	&	-3.8985	\\
0.03	&	-5.5454	&	0.13	&	-4.6807	&	0.23	&	-4.1776	&	0.33	&	-3.8791	\\
0.04	&	-5.4206	&	0.14	&	-4.6200	&	0.24	&	-4.1385	&	0.34	&	-3.8627	\\
0.05	&	-5.3106	&	0.15	&	-4.5619	&	0.25	&	-4.1013	&	0.35	&	-3.8495	\\
0.06	&	-5.2112	&	0.16	&	-4.5064	&	0.26	&	-4.0660	&	0.36	&	-3.8399	\\
0.07	&	-5.1199	&	0.17	&	-4.4532	&	0.27	&	-4.0327	&	0.37	&	-3.8343	\\
0.08	&	-5.0353	&	0.18	&	-4.4022	&	0.28	&	-4.0014	&	0.38	&	-3.8332	\\
0.09	&	-4.9561	&	0.19	&	-4.3533	&	0.29	&	-3.9722	&	0.39	&	-3.8678	\\
0.1	&	-4.8816	&	0.2	&	-4.3065	&	0.3	&	-3.9453	&	0.4	&	*******	\\
%\vspace{0.1cm}
 \hline  
    \end{tabular}
  \label{t1}
\end{table}

\begin{figure}[H]
%\isPreprints{\centering}{} % Only used for preprints
\begin{center}
\includegraphics[width=0.7\textwidth]{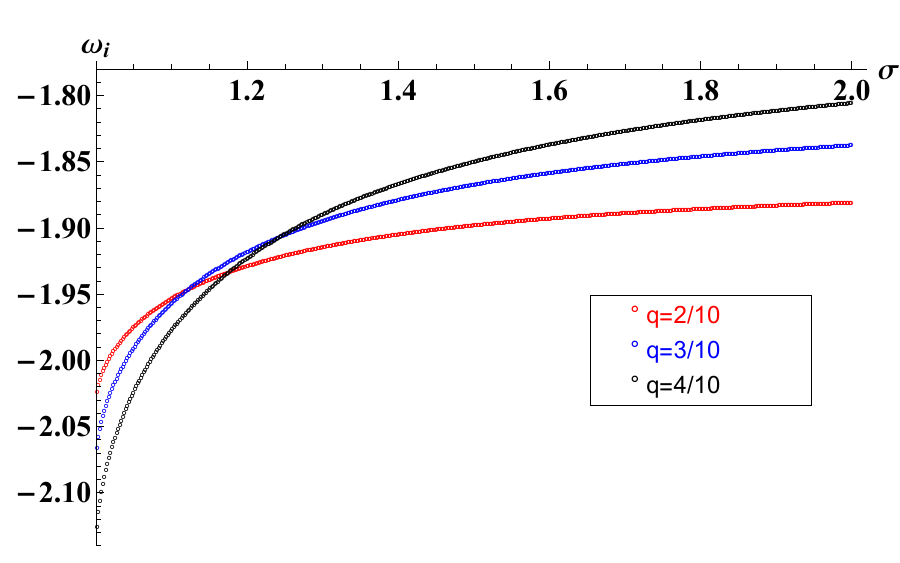}
\end{center}
\caption{The fundamental quasinormal modes of black holes in nonlinear electrodynamics. All the solutions are purely imaginary with no angular momentum ($k=0$). The geometry parameters are $M=\mathfrak{L}=1$.}
\label{fig5}
\end{figure}

The last results we present for the scalar field perturbations are those for the black holes with  one horizon and $\sigma >1$. In such scope ($0<\Sigma < 1$) we investigate the response that different theories bring for a scalar field perturbations with fixed charges. The results are displayed in figure \ref{fig5}.
%Cargas e sigma assintótico.

%\begin{adjustwidth}{-\extralength}{0cm}
%\end{adjustwidth}

In the figure we can observe the opposite relation between the increase in $\sigma$ and the decrease of $|\omega_i|$ approaching asymptotically to a fixed value, when $\Sigma = 1$. We calculate asymptotic quasinormal modes $\omega_i$ with different charges in the limit $\Sigma = 1$ and the results are listed in table \ref{t2}.

%\begin{table}[h]
% \centering
%\tiny
%\begin{adjustwidth}{-\extralength}{1cm}
%\addtolength\tabcolsep{6pt}
%\setlength{\tabcolsep}{10.5pt} % Default value: 6pt
%\renewcommand{\arraystretch}{1.5} % Default value: 1
%   \begin{tabularx}{\fulllength}{c|cccccccccc}
%\vspace{0.2cm}
%   \begin{tabular}{c|cccccccccc}
%    \hline    \hline
%\vspace{0.1cm}
%$q$	&	1/2	&	1	&	3/2	&	2	&	5/2	&	3	&	7/2	&	4	&	9/2	&	5	\\
%$-\omega_i$	&	1.6918	&	1.6470	&	2.0816	&	3.0235	&	4.4017	&	6.1651	&	8.2886	&	10.760	&	13.574	&	16.727	\\
% \hline  
%    \end{tabularx}
%\end{tabular}
%    \end{adjustwidth}
%  \label{t2}
% \caption{\color{black} Asymptotic quasinormal modes for theories with $\Sigma \rightarrow 1$ and zero angular momentum for the scalar field. The geometry parameters read $M=\mathfrak{L}%=1$.}
%\end{table}
%\end{adjustbox}
%

\begin{table}[h]
\centering
 \caption{\color{black} Asymptotic quasinormal modes for theories with $\Sigma \rightarrow 1$ and zero angular momentum for the scalar field. The geometry parameters read $M=\mathfrak{L}=1$.}
\small
\begin{tabular}{c|cccccccccc}
 \hline    \hline
\vspace{0.1cm}
$q$	&	1/2	&	1	&	3/2	&	2	&	5/2	&	3	&	7/2	&	4	&	9/2	&	5	\\
$-\omega_i$	&	1.6918	&	1.6470	&	2.0816	&	3.0235	&	4.4017	&	6.1651	&	8.2886	&	10.760	&	13.574	&	16.727	\\
 \hline  
\end{tabular}
\label{t2}
\end{table}

In the particular cases shown in figure \ref{fig5} we have  $\omega_i=-1.86222i$, $\omega_i=-1.7985$ and $\omega_i=-1.7410i$ for the asymptotic oscillations for the charges $q=1/5$,
$q=3/10$ and $q=2/5$ respectively.

It is important to notice that all field dynamics studied along this section points in the direction of a stable spacetime to first order scalar perturbations, as long as no frequencies were found with a positive coefficient $\omega_i$ representing unstable evolutions. Although this fact does not constitute a mathematical proof of geometric stability to first order perturbations, it strongly suggests that the black holes of such theories react to a massless and changeless scalar field with a tower of quasinormal oscillations decaying in time, thus maintaining the event horizon of the black hole and the fabric of spacetime intact.

\section{Final remarks}\label{sec5}

In this work we studied black holes in 2+1 nonlinear electrodynamic theories.

The geometry of different line-elements representing possible solutions was studied, particularly in relation to the presence of horizons. We found a rich causal structure depending on the $\sigma$-coefficient considered. Interestingly enough the spacetime is regular at $r=0$ only for small $\sigma$´s.

In the case of $\sigma < 1/2$ we have black holes regular at $r=0$ with one event horizon, in all possible solutions with charge, mass and cosmological constant. That is clearly not the case of the linear theories either in 2+1 or in higher dimensional spacetimes. 

Intermediate values of $\sigma$ in the scope $(-1/2,1)$ bring solutions with two or no horizons depending on the values of the geometry constants. As common to other charged spacetimes in such cases we have a maximum value of charge, here calculated through Eq. (\ref{e17}), depending on $M,\Lambda$ and $\sigma$ as of a naked curvature singularity is present at $r=0$.

For high values of $\sigma$ $(>1)$, we also have one horizon present for every charge, mass and cosmological constant. The behavior is essentially distinct as that presented in the intermediate values (and linear theories) by the change of s in (\ref{e9}). In such cases, we have a curvature singularity at $r=0$ hidden by the event horizon. 

We also studied the propagation of a scalar field delivered via matter action in all three types of theories. Our results strongly suggest the spacetimes to be stable to first order perturbations with the scalar field evolution being given in terms of the typical quasinormal evolution. The quasinormal spectra of the  different theories were calculated in section \ref{sec4}. They show an intriguing peculiarity of diminishing $|\omega_i|$ as $\sigma$ approaches $1/2$. In the cases of asymptotic $\sigma$, $\Sigma \rightarrow 1$ and for high charges $\omega_i$ increases unboundedly with $q$. Interesting enough, that behavior points in the opposite direction of the interpretation of $\omega_i$ proportional to the Hawking temperature of the hole \cite{hh2000} thus not representing the relaxation time of the lower dimensional conformal field theory: in such cases $T_H = r_+/2\pi \mathfrak{L}^2$ \cite{hendi14} decreasing with the increment of $q$. \textcolor{black}{Such interpretation is also lost in the near extremal cases of the linear theory of the charged BTZ black hole (see e. g. \cite{Fontana_2024}), although for intermediate and small values it endures.}

Further lines of investigation include the inspection of superradiant phenomena that may be associated with a charged scalar propagation and the perturbations delivered by fields of different spins. \textcolor{black}{Also interesting is the examination of rotating black holes with non-linear electrodynamic terms as well as that with acceleration \cite{rincon25}, as brought by one of our anonymous referees.}

\acknowledgments{
This work was partially supported by CNPq (Conselho Nacional de Desenvolvimento Científico - Brazil) under Grant No. $405749/2023-6$. The authors would like to thank J. S. E. Portela and J. de Oliveira for fruitful discussions.}

%\conflictsofinterest{The authors declare no conflicts of interest.} 

\bibliography{references}

\end{document}